\documentclass[epj,twocolumn, 12pt]{webofc}
\usepackage[varg]{txfonts}   % Web of Conferences font
\wocname{EPJ Web of Conferences}
\woctitle{XLIV International Symposium on Multiparticle Dynamics 2014}

\DeclareGraphicsExtensions{.eps}
\usepackage{amsmath}
\usepackage{comment}
\usepackage{geometry}

\newcommand{\tch}{T_{ch}}
\newcommand{\mub}{\mu_{B}}
\newcommand{\muq}{\mu_{Q}}
\newcommand{\mus}{\mu_{S}}
\newcommand{\gs}{\gamma_{S}}

\newcommand{\snn}{\sqrt{s_{NN}}}
\newcommand{\np}{\langle N_{\textnormal{part}} \rangle}

\usepackage[colorlinks=true, pdfstartview=FitH, linkcolor=blue, citecolor=blue, urlcolor=blue, bookmarksopen=true]{hyperref}
\begin{document}
\title{Chemical freeze-out parameters in Beam Energy Scan Program of STAR at RHIC}

\author{Sabita Das~(for the STAR collaboration)\inst{1}\fnsep\thanks{\email{sabita@rcf.rhic.bnl.gov}}
}

\institute{Institute of Physics, Bhubaneswar-751005, India}

\abstract{%
The STAR experiment at RHIC has completed its first phase of the Beam Energy Scan~(BES-I)
program to understand the phase structure of the quantum
chromodynamics~(QCD). The bulk properties of the system formed in
Au+Au collisions at different center of mass energy 
$\snn = $ 7.7,
11.5, 19.6, 27, and 39 GeV have been studied from the data collected in the year 2010 and 2011. 
The centrality and energy
dependence of mid-rapidity~($|y|$ < 0.1) particle
yields, and ratios are presented here. The chemical freeze-out
parameters are extracted using measured particle ratios within the
framework of a statistical
model. 
}
\maketitle
\section{Introduction}
\label{intro}
The heavy-ion collider experiments such as STAR at RHIC, ALICE
at LHC were designed to investigate matter similar to that formed at the very early stages of the
universe {\it i.e.} matter under extreme conditions of
high temperature or density~(or both)~\cite{qgp1}. Similar to the
matter in its primordial state, a deconfined state of quarks and gluons is created called Quark-Gluon
Plasma~(QGP) at both RHIC and LHC~\cite{qgprhic,
  qgplhc}. The QCD 
as the theory of strong interactions predicts a transition at sufficiently high
temperature $T$ or baryon chemical potential $\mub$ from hadronic matter to QGP state. 
So, by varying the $T$  and $\mub$ in laboratory we can study the
phase transition associated with QCD matter~\cite{science, bb1, bb2, bb3}. The major part of the 
QCD phase diagram, which is generally known as the plot of $T$  as a
function of $\mub$, consists of two phases ~\cite{tmubth}. These are
the high temperature QGP phase, where
the relevant degrees of freedom are quarks and gluons, and the hadronic phase at low temperature. Other interesting phases
related to neutron stars~\cite{neutronstar5}, color
superconductivity~\cite{superconductivity6}, and the quarkyonia~\cite{quarkonia} phase also appear in the QCD phase diagram in addition to the confined and de-confined
phases~\cite{qcdphasedgm}. Particle yields in high energy heavy-ion
experiments at different collision energies can be used to obtain the
$T$ and $\mu_{B}$ that set up the chemical freeze-out line in the QCD phase diagram. It appears to be very close to the
phase boundary between QGP and hadronic phase, especially at low
$\mub$. At high $T$ and vanishing $\mub$, finite temperature lattice QCD
calculations has established the transition from QGP to a hadron gas 
is a cross-over~\cite{nature6}. The existence of a first-order phase
transition has been predicted by several QCD-based
 calculations at lower $T$ and $\mu_{B}$~\cite{1st}. The QCD critical point is a feature of
the phase diagram, where the nature of the transition changes from a discontinuous (first-order)
transition to an analytic crossover~\cite{qcd1, qcd2, lqcd1,lqcd2, lqcd3}.\\ 
%-------------------------------------------
\begin{figure*}[!hbtb]
\begin{center}
\includegraphics[scale=0.43]{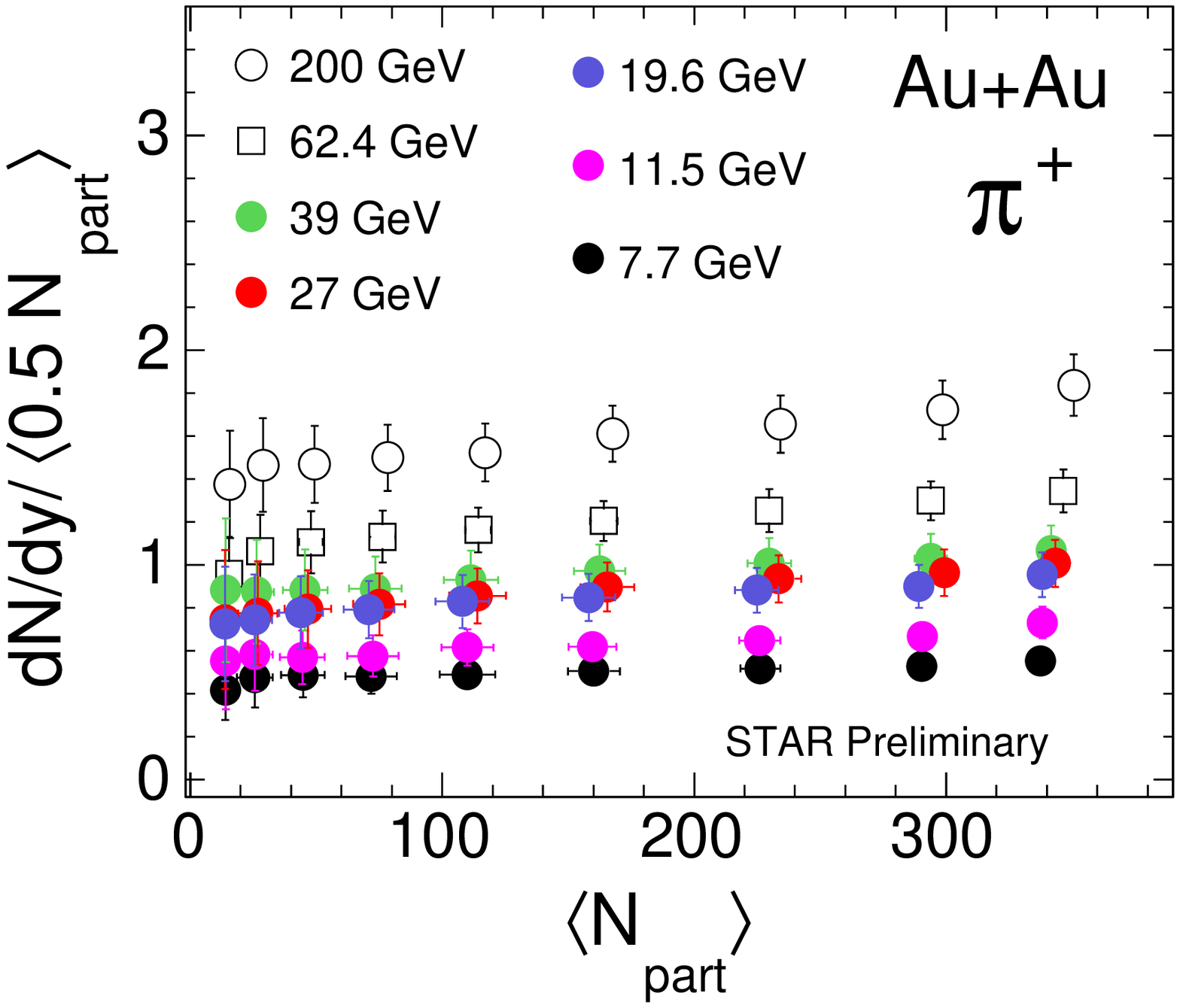}
\hspace{-0.66cm}
\includegraphics[scale=0.43]{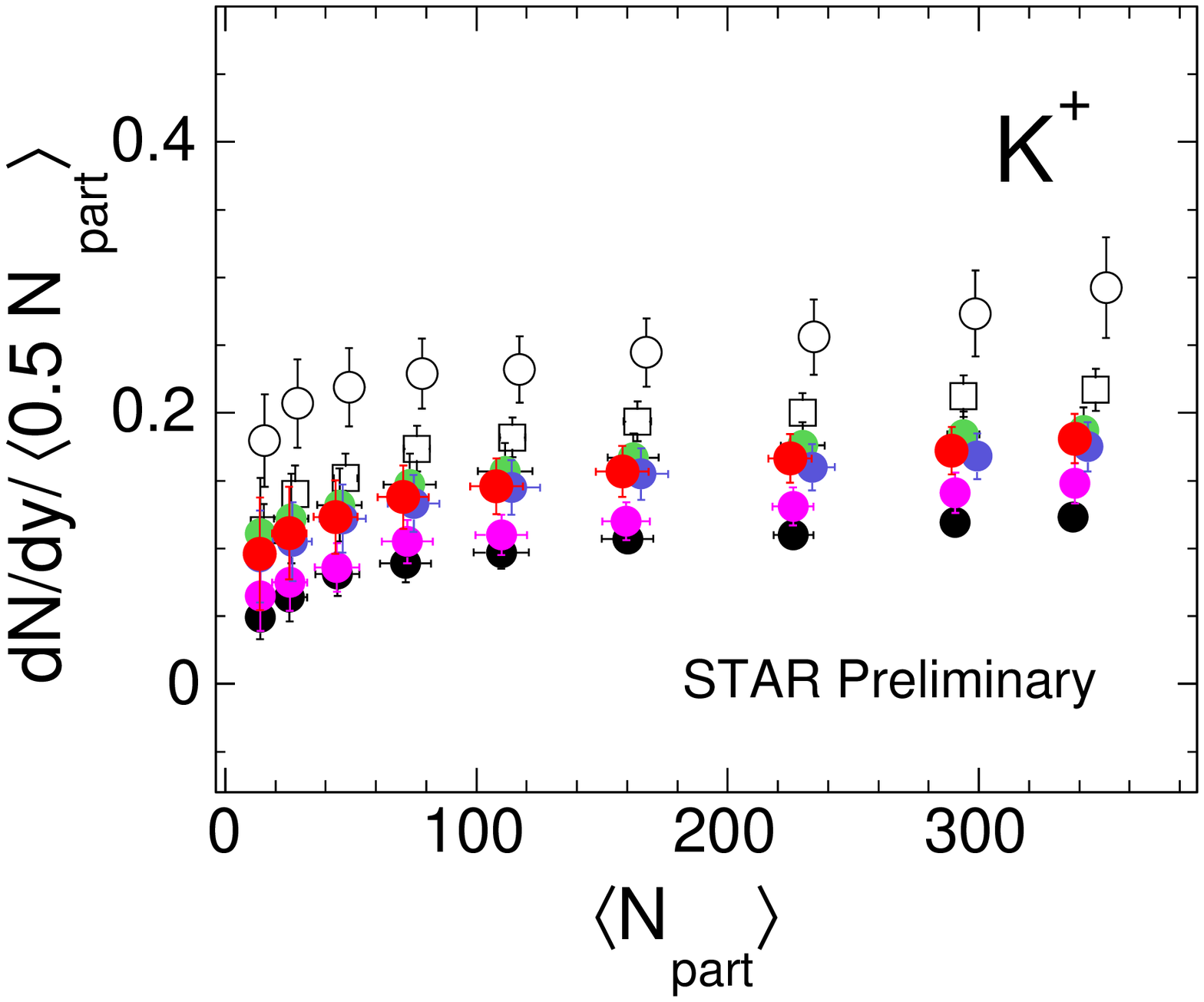}
\hspace{-0.66cm}
\includegraphics[scale=0.43]{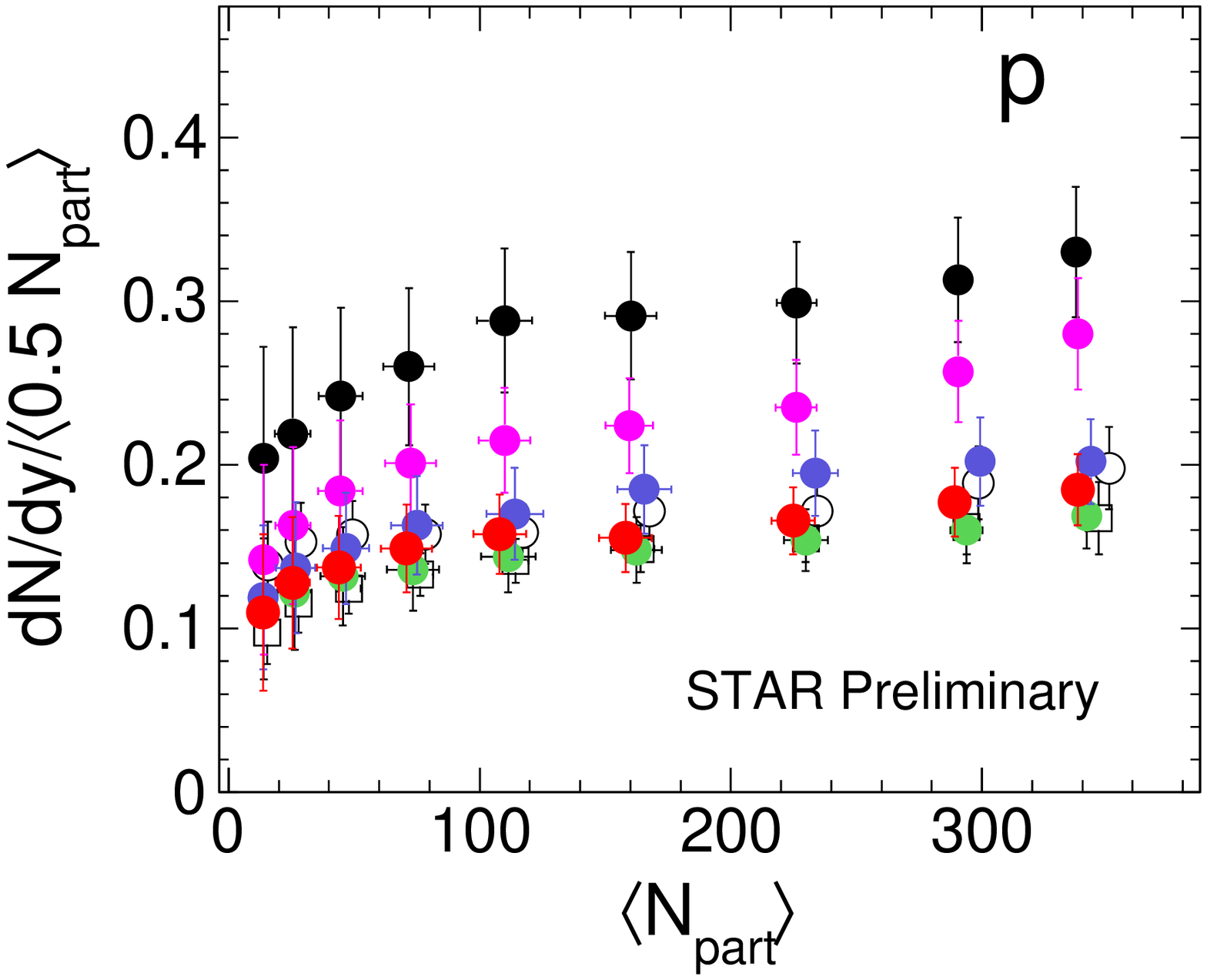}
\caption{$dN/dy$ of $\pi^{\pm}$, $\emph{K}^{\pm}$, and $\emph{p}$ scaled by $\langle 0.5N_{part} \rangle$ as a
function of $\langle N_{part} \rangle$ in Au+Au
collisions at BES energies~(solid symbols) along with top RHIC
energies~(open symbols)~\cite{datastar}. 
Errors are statistical and systematic errors added in quadrature.}
\label{fCentdndy}
\end{center}
\end{figure*}
%------------------------------------------
To explore the freeze-out diagram, {\it i.e.} search
the possible phase boundary line and search for the possible QCD critical
point, is a priority study at RHIC. For this, STAR has
completed the first phase of the Beam Energy Scan (BES-I)
program~\cite{starnote2009}, collecting data from Au+Au collisions at
center of mass energies of 7.7, 11.5, 14.5, 19.6, 27, and 39 GeV in the
year 2010, 2011 and 2014. In this paper, we use the particle yields or ratios of the
measured hadrons to find the experimentally accessible region of the QCD phase
diagram by extracting
$T, ~\mub$ within a statistical model.\\   
Here we will discuss the centrality dependence of identified particles,
~pions ($\pi^{\pm}$), ~kaons ($\emph{K}^{\pm}$), ~protons ($\emph{p}$),
and ~anti-protons~($\bar{\emph{p}}$) produced
in Au+Au collisions at BES energies $\sqrt {s_{NN} }=
7.7$, 11.5, 19.6, 27, and 39 GeV~\cite{sd, lk}. 
The experimental particle ratios obtained from the yields of
$\pi^{\pm}$, $\emph{K}^{\pm}$, $\emph{p}$,
$\bar{\emph{p}}$, $\Lambda$, $\bar{\Lambda}$
and $\Xi^{-}$, $\bar{\Xi}^{+}$ ~\cite{zhu} have been used in a grand-canonical
ensemble~(GCE) of THERMUS model~\cite{thms} to extract the freeze-out parameters like 
chemical freeze-out temperature ($T_{ch}$),
baryon chemical potential ($ \mu_{B}$), strangeness chemical potential
($ \mu_{S}$), charge chemical potential ($ \mu_{Q}$), and strangeness saturation factor ($\gamma_{S}$). The
centrality and energy dependence of these parameters will also be
discussed.
\section{Results}
%-------------------------------------------
\begin{figure*}[!hbtb]
\begin{center}
\includegraphics[scale=0.42]{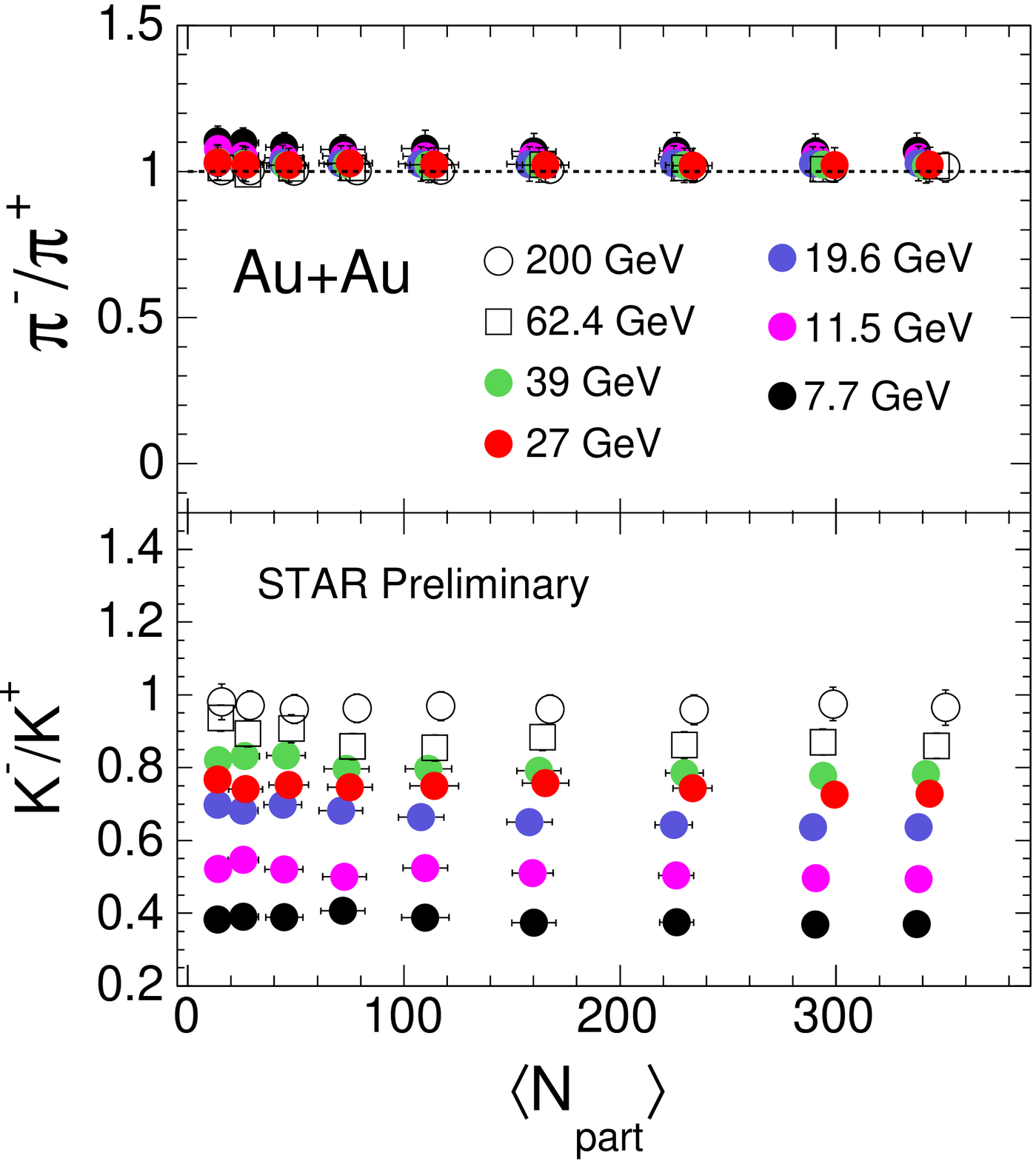}
\includegraphics[scale=0.42]{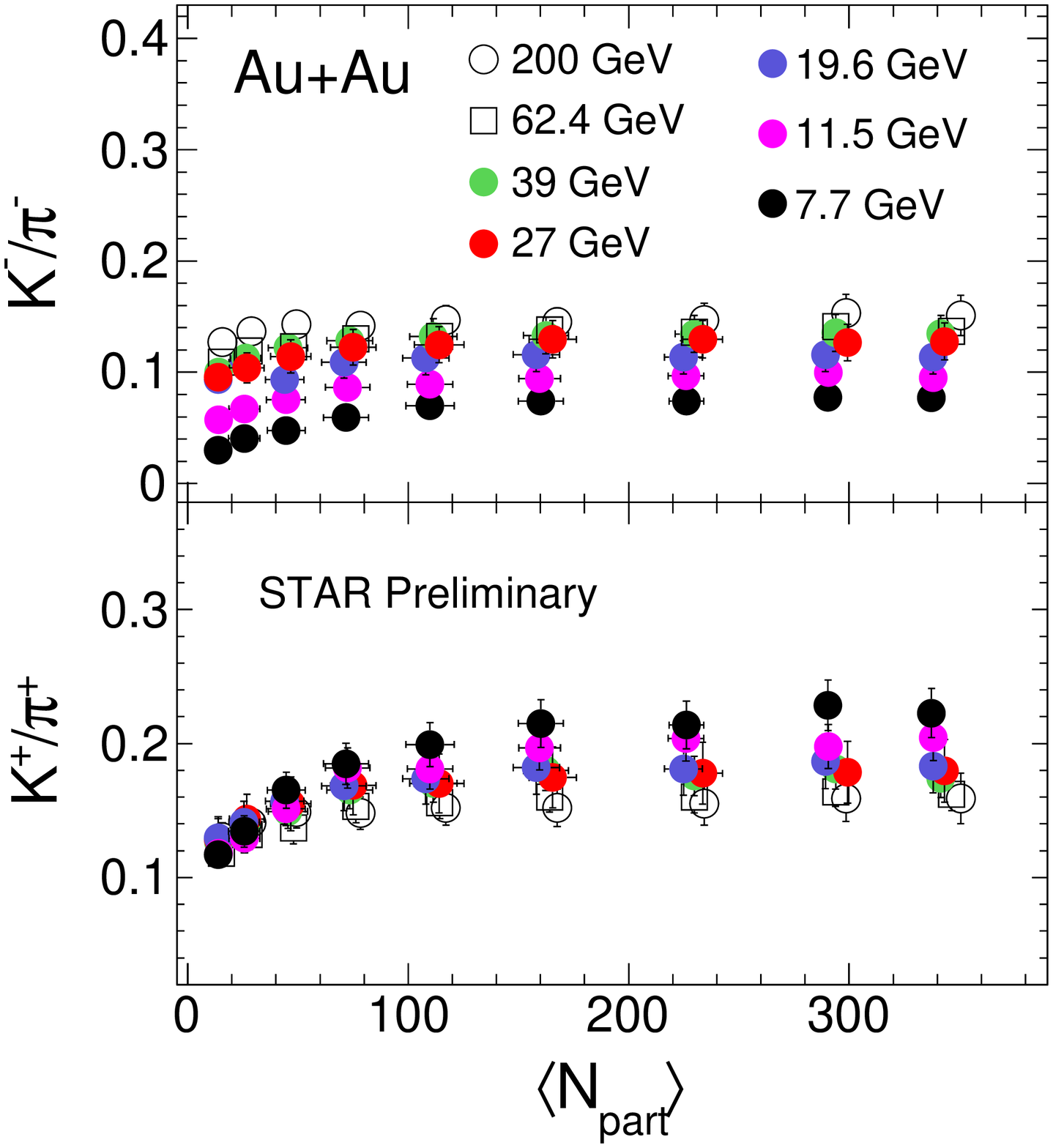}
%\includegraphics[scale=0.23,height = 6.5cm,width =6cm]{./combine_ratio_kappip_kampim.eps}
%\hspace{-0.65cm}
%\includegraphics[scale=0.25]{./ismd_kmkp_vsprmprp_vsalleng.eps}
%\includegraphics[scale=0.37,height = 6.5cm,width =7.35cm]{./figCh4/chk_muq/condn3_mus_npart_gce_yld_7gev.eps}
\caption{$\pi^{-}/\pi^{+}$, $\emph{K}^{-}/\emph{K}^{+}$,
  $\emph{K}^{-}/\pi^{-}$, and $\emph{K}^{+}/\pi^{+}$ as a function of $\langle N_{part} \rangle$ in Au+Au
 collisions at BES energies and top RHIC energies~\cite{datastar}. Errors are statistical and systematic errors added in quadrature.}
\label{fCratios}
\end{center}
\end{figure*}
%-------------------------------------------
\subsection{Particle Yields}
The identified particle yields are measured using the STAR Time Projection chamber
(TPC) and Time-Of-Flight (TOF) detectors in mid-rapidity ($|y| <
0.1$)~\cite{tpctof}. In the TPC, the particles are identified by measuring
the specific ionisation energy loss whereas in TOF, they are identified using the particle
velocities as a function of momentum. 
Figure~\ref{fCentdndy} shows the $dN/dy$ normalised to the average number of participated nucleon pair
($dN/dy/ \langle 0.5N_{part} \rangle $) vs. $\langle N_{part} \rangle$
for $\pi^{+}$, $\emph{K}^{+}$, and
$\emph{p}$ where $\np$ is the average number of participating nucleons
in a given centrality range. The errors on the points which include
both statistical and systematic errors. For the comparison, the $dN/dy/ \langle 0.5N_{part} \rangle $ from the Au+Au system
at RHIC energies are also shown~\cite{datastar}.%%check. 
~The yield per participant pair for pions at BES energies is almost independent of
$\langle N_{part} \rangle$. The particle production of pions therefore scales
with the number of participant pairs. Kaon yields
per participant pair increases with average participant number. The increase in proton yield
per participating nucleon with the increasing collision centrality is
possibly due to large baryon
stopping at the lower energies. 
\subsection{Particle Ratios}
Figure~\ref{fCratios} show the centrality
dependence of different particle ratios in Au+Au
 collisions at BES energies and its comparison 
 with other RHIC energies~\cite{datastar}. The errors shown are the
 quadratic sum of statistical and systematic errors. 
The $\pi^{-}/\pi^{+}$ ratio shows no strong dependence
on centrality. 
The flat $\emph{K}^{-}/\emph{K}^{+}$ ratios vs. $\np$ indicate that the production mechanism does not change across
centrality.\\
The ratios of $\emph{K}^{+}/\pi^{+}$ and $\emph{K}^{-}/\pi^{-}$ gradually increase
from peripheral to mid-central and saturate in mid-central to central
collisions. The enhancement of the integrated $\emph{K}^{\pm}/\pi^{\pm}$ ratio in more central collisions is related to strangeness
equilibration in various thermal models~\cite{smodelkpi1, smodelkpi2}. Thermodynamic models explain the increase of the $\emph{K}/\pi$
ratios with the system size from peripheral to central based on the transition from the canonical to grand-canonical
ensemble ~\cite{kpi1, kpi2}.
%-------------------------------------------
\begin{figure*}[!hbtb]
\begin{center}
\hspace{0.2cm}
\includegraphics[scale=0.4]{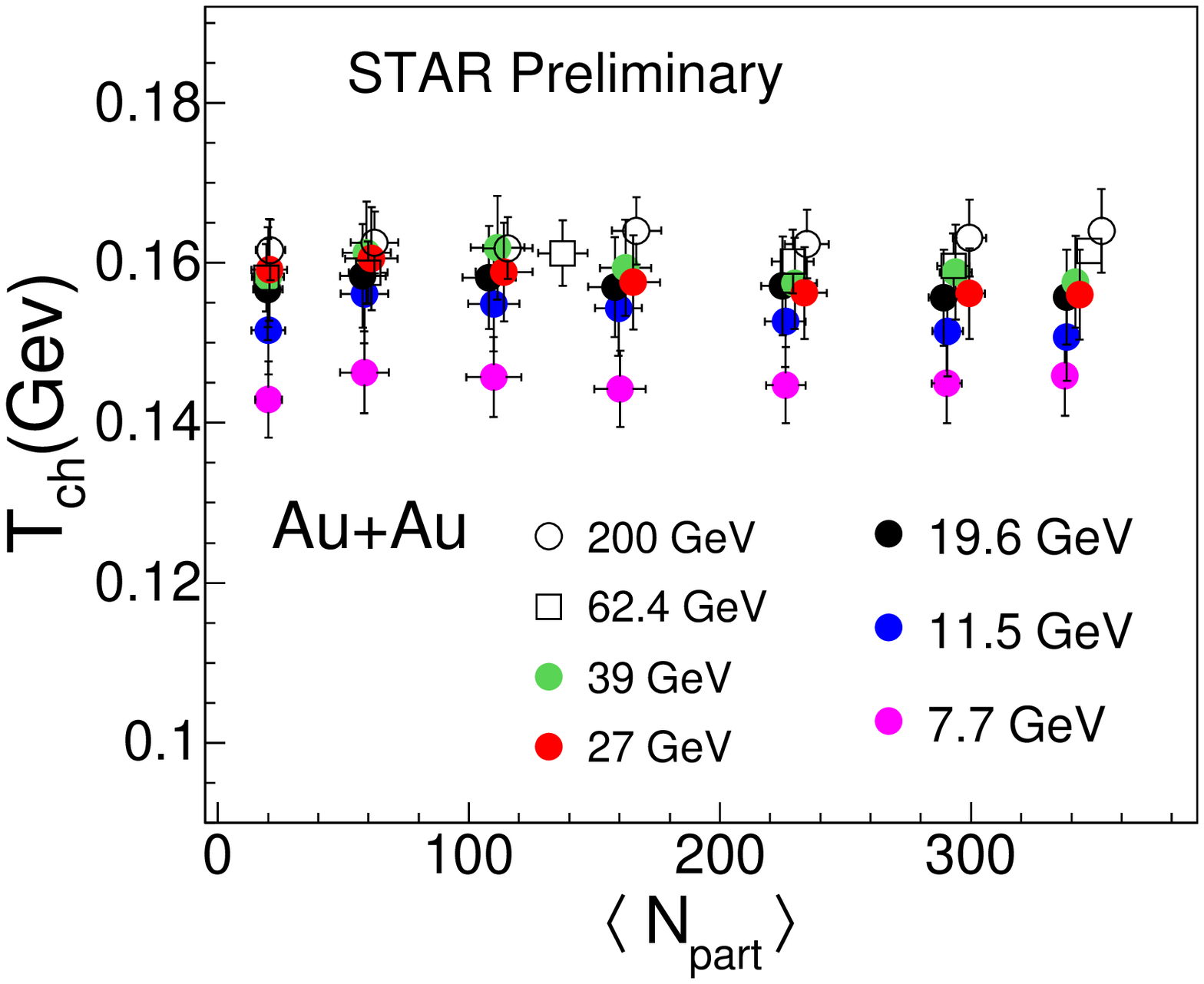}
\hspace{0.3cm}
\includegraphics[scale=0.4]{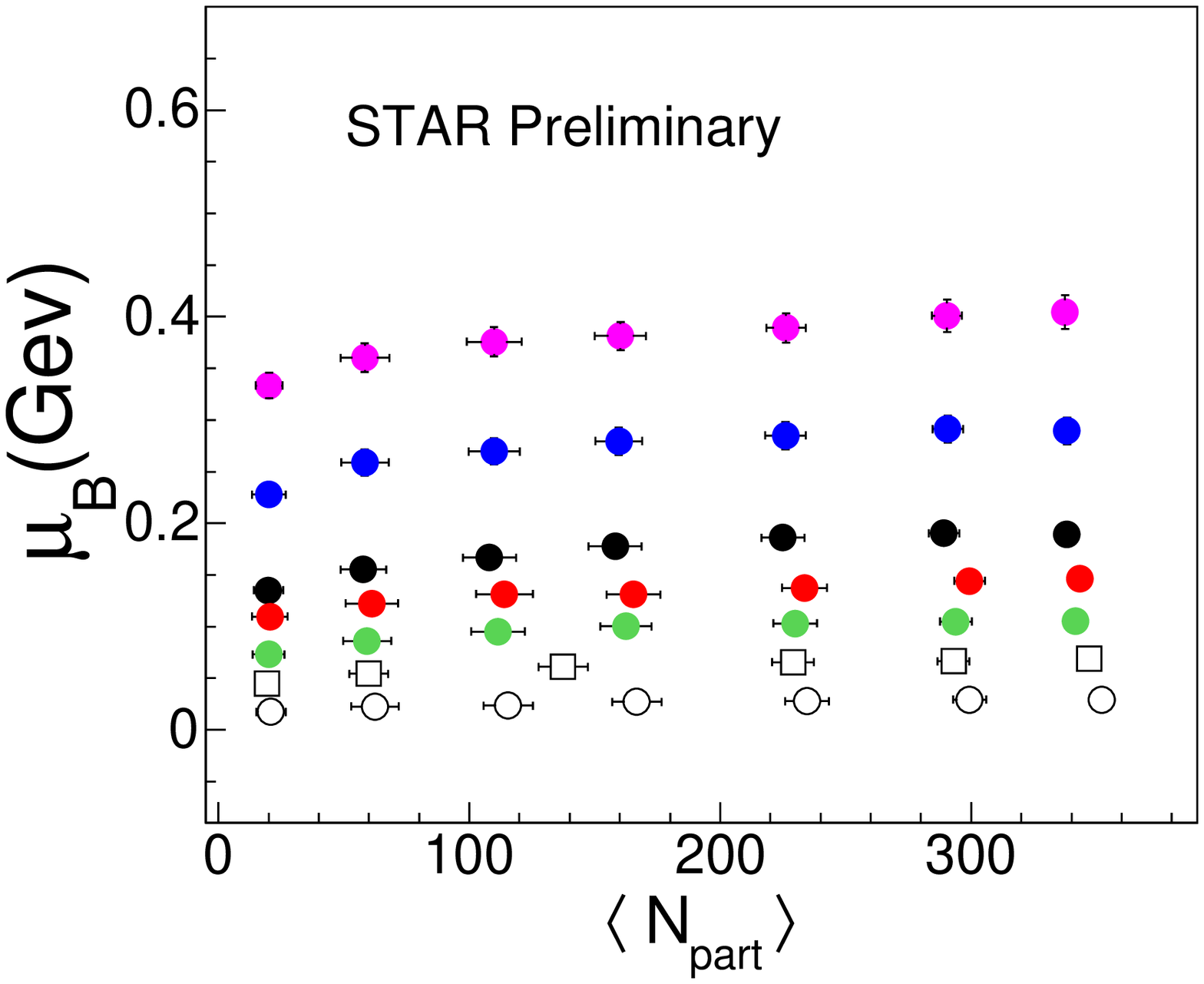}
\hspace{0.2cm}
\includegraphics[scale=0.4]{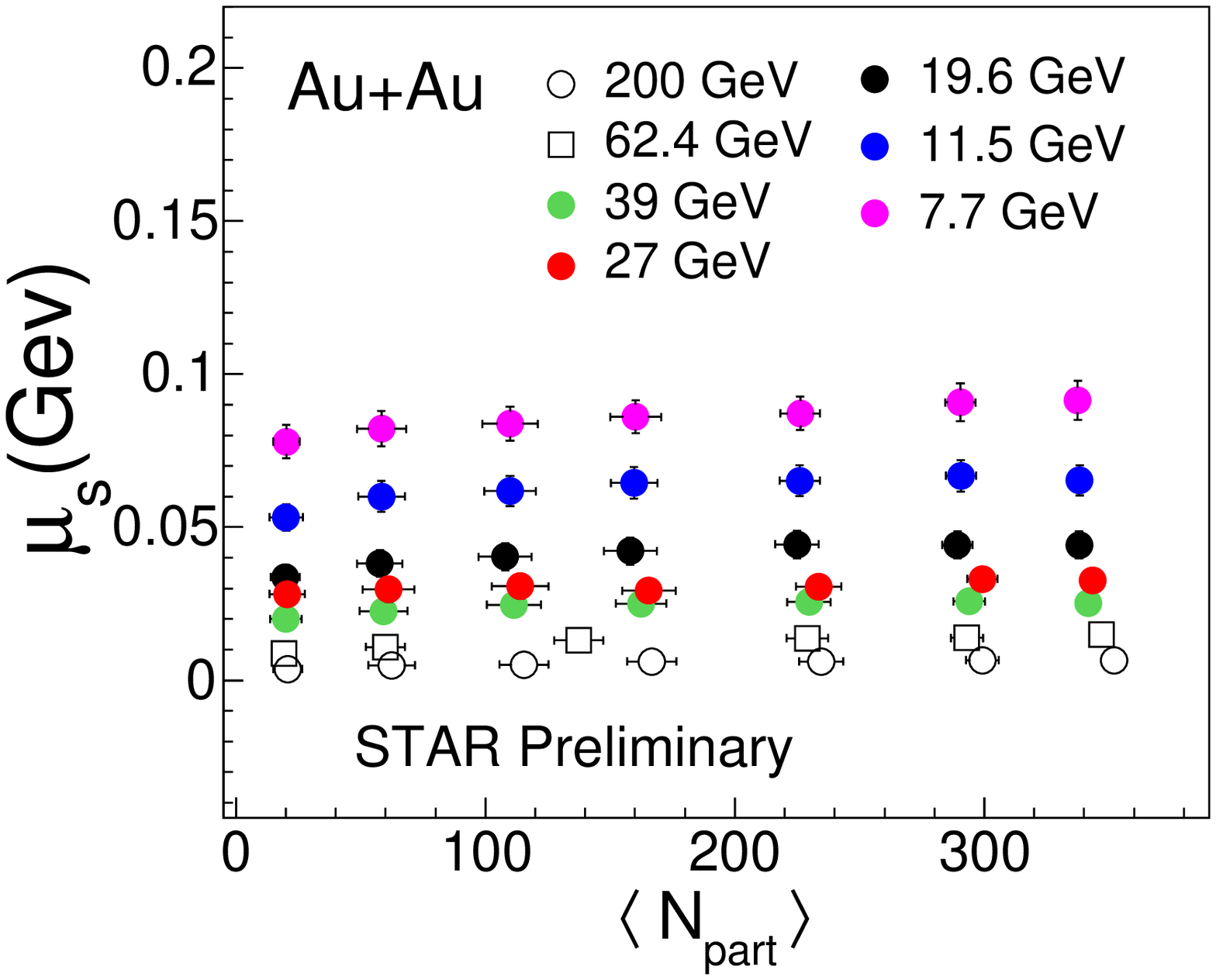}
\hspace{0.3cm}
\includegraphics[scale=0.4]{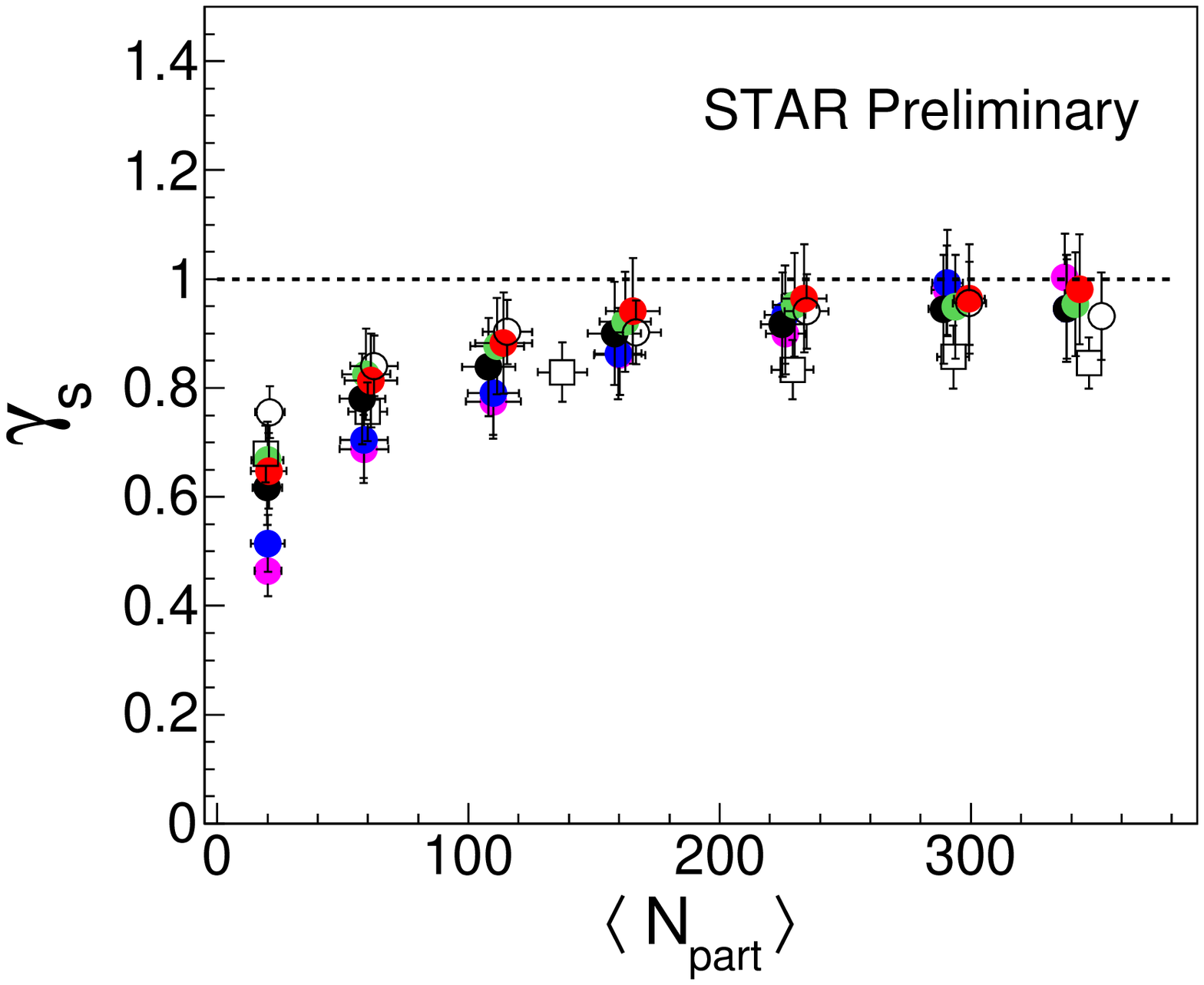}
%\includegraphics[scale=0.32]{./ismd_thesis_mub_alleng_sce.eps}
%////////need to add later
\caption{Chemical freeze-out
 temperature $\tch$, baryon chemical potential $\mub$, strangeness
 chemical potential $\mus$, strangeness saturation factor $\gs$ are shown as a function of the average number of
 participating nucleons $\np$ in Au+Au
collisions at $\sqrt{s_{NN}} = 7.7$, 11.5, 19.6, 27, 39, 62.4, and 200
GeV obtained from particle ratios in grand-canonical ensemble.}
\label{frgce}
\end{center}
\end{figure*}
%-------------------------------------------
\subsection{Chemical freeze-out}

The scenario when inelastic collisions among the particles stop
and particle ratios get fixed, is called the chemical
freeze-out. We have chosen to use THERMUS~\cite{thms}
model assuming chemical equilibrium to study chemical freeze-out dynamics for BES energies.
The chemical freeze-out parameters, $\tch,~\mub,~\mus$, and ~$\gs$,
are extracted from the mid-rapidity particle ratios of different
combinations which includes yields of $\pi^{\pm}$, $\emph{K}^{\pm}$, $\emph{p}$,
$\bar{\emph{p}}$, $\Lambda$, $\bar{\Lambda}$,
$\Xi^{-}$, $\bar{\Xi}^{+}$~\cite{sd, lk, zhu}. Experimentally, the proton yields have not
been corrected for feed-down contributions. The yields of $\pi$ and
$\Lambda$ have been corrected for the feed-down
from weak decays.\\ 
Figure ~\ref{frgce} shows the centrality and energy dependence of
freeze-out parameters as a function of average number of participants $\np$, 
using THERMUS with GCE for Au+Au collisions at $\snn = $ 7.7 - 200 GeV. The solid
symbol represents the results for BES energies and the open symbol
represents the results for top RHIC energies. Here we always use $\muq = 0$
with unconstrained fit parameters $\tch,~\mub,~\mus$, and ~$\gs$. The $T_{ch}$ is found to be independent of
centrality and with increasing energy its value increases from lower
energy 7.7 GeV up to 19.6 GeV and afterwards its value remain consistent within error with the top RHIC
energy measurements. The values of $\tch$ is close to the lattice
QCD calculation of the cross-over temperature between the
de-confined phase and the hadronic phase suggesting
that chemical freeze-out happens in the vicinity of the
transition line shortly after hadronization. The baryon chemical
potential, $\mu_{B}$, decreases with increasing collision energy and it
increases from peripheral to central collisions at all
energies. We observe a centrality dependence of the chemical freeze-out
curve (T$_{ch}$ vs. $\mu_{B}$) at BES energies
which was not observed at higher energies of Au+Au 200
GeV~\cite{sd, datastar}. The strangeness chemical potential, $\mus$, 
seems to decrease with the increasing collision energy, following the
same type of behaviour as $\mub$. The strangeness saturation factor
$\gs$ increases from peripheral to central for all the
energies studied. In central collisions, $\gs$ is close to unity for top
RHIC energies.
\subsection{Summary}
The centrality and energy dependence of identified particle yields and ratios have been discussed in Au+Au
collisions at BES energies $\sqrt{s_{NN}} = 7.7$, 11.5, 19.6, 27, 39 GeV. The yields of identified hadrons
increase with increase of beam energy. The freeze-out
temperature and baryon chemical potential can be deduced from
statistical fits of experimentally measured
particle ratios to thermal model calculations assuming chemical
equilibrium. 
The measurements at BES energies along with the top RHIC energies 
have extended the range of baryon
chemical potential ($\mu_B$) from 20 to 420 MeV corresponding to Au+Au
collision at $\sqrt {s_{NN} }=$ 200 down to 7.7 GeV. A centrality
dependence of freeze-out parameters~($\tch,~\mub$) has been observed at BES energies.

% Non-BibTeX users please use

\end{document}